\documentclass[showpacs,amsmath,amssymb,aps,pra,preprint,groupedaddress]{revtex4-1}

\usepackage{graphicx}
\usepackage[latin1]{inputenc}

\begin{document}

\newtheorem{theorem}{Teorema}[section]
\newcommand{\n}[1]{{\bf #1}}

%\preprint{}

\title{On the accelerated observer's proper coordinates and the rigid motion problem in Minkowski spacetime}


\author{J. B. Formiga}
\email[]{jansen.formiga@uespi.br}
%\thanks{}
\affiliation{Centro de Ciências da Natureza, Universidade Estadual do Piauí,  64002-150 Teresina, Piauí, Brazil}


\date{\today}

\begin{abstract}
Physicists have been interested in accelerated observers for quite some time. Since the advent of special relativity, many authors have tried to understand these observers in the framework of Minkowski spacetime.  One of the most important issues related to these observers is the problematic definition of rigid motion. In this paper, I write the metric in terms of the Frenet-Serret curvatures and the proper coordinate system of a general accelerated observer. Then, I use this approach to create a systematic way to construct a rigid motion in Minkowski spacetime. Finally, I exemplify the benefits of this procedure by applying it to two well-known observers, namely, the Rindler and the rotating ones, and also by creating a set of observers that, perhaps, may be interpreted as a rigid cylinder  which rotates while accelerating along the axis of rotation.
\end{abstract}

%\keywords{Special relativity; accelerated observers; rigid motion.}

\maketitle

\section{Introduction}
Accelerated observers in Minkowski spacetime have been widely studied in physics and there is no doubt about their importance to modern physics. They have been used to study quantum phenomena, like the  Unruh effect \cite{Unruh:1976db}, and to understand some properties of general relativity \cite{Klauber:2006wv,Klauber:1998rm}. Some very nice papers on the subject have been published so far \cite{Unruh:1976db,Klauber:2006wv,Klauber:1998rm,Rosen1947,Mashhoon:2009tz,Mashhoon:2005fe,Mashhoon:2003st,Mashhoon:2003pg,Mashhoon:2002xs,Formiga:2006ur,Maluf:2010fb,Maluf:2011gt}, some of them trying to answer fundamental questions such as ``how do electric charges behave in an accelerated frame?'' \cite{Maluf:2010fb,Maluf:2011gt}. Another important use of these observers lies in the definition of rigid motion in Minkowski spacetime, which is a controversial issue. As an example of the important role played by accelerated observers, we have the rotating observers, which are generally used to deal with a rigid disk \cite{Rosen1947}.

In Sec. \ref{21102012s2} of this paper, I use the tetrad formalism to obtain an expression for the metric tensor in terms of the proper coordinate system of an arbitrary accelerated observer. I also write the metric tensor in terms of the curvatures of the observer's curve. I use this approach and the definition of rigid motion presented in Ref. \cite{Rosen1947} to create a systematic way to construct a rigid motion in Minkowski spacetime. To exemplify the benefits of using this approach, in Sec. \ref{21102012s3}, I apply it to the Rindler  and rotating observers. In addition, I create a new set of  observers that perhaps can be used to represent a particular motion of a rigid cylinder. A brief introduction to the Frenet-Serret tetrad is given in Sec. \ref{21102012s1}.

Throughout this paper capital Latin letters represent tetrad indices, which run over (0)-(3), while the Greek ones represent coordinate indices, which run over 0-3; the small Latin letters run over 1-3. The frame is denoted by $e_A$, and its components in the coordinate basis $\partial_{\mu}$ are represented by $e_A^{\ \ \mu}$.

\section{Frenet-Serret Tetrad \label{21102012s1}}
Let $x^{\mu}(s)$ be a curve in Minkowski spacetime, where $s$ is its arc length. In this spacetime, Frenet-Serret basis can be defined through the formulas
\begin{eqnarray}
\frac{d e_{(0)}^{\ \ \mu}}{ds}=k_1 e_{(1)}^{\ \ \mu}, \label{22102012a}\\
\frac{d e_{(1)}^{\ \ \mu}}{ds}=k_1 e_{(0)}^{\ \ \mu}+k_2 e_{(2)}^{\ \ \mu}, \label{22102012b}\\
\frac{d e_{(2)}^{\ \ \mu}}{ds}=k_3 e_{(3)}^{\ \ \mu}-k_2 e_{(1)}^{\ \ \mu}, \label{22102012c}\\
\frac{d e_{(3)}^{\ \ \mu}}{ds}=-k_3 e_{(2)}^{\ \ \mu}, \label{22102012d}
\end{eqnarray}
where $e_{(0)}^{\ \ \mu}=d x^{\mu}/ds$ and $e_A^{\ \ \mu}$ are the components of the vectors in the Cartesian coordinate basis $\partial_{\mu}$ (for a general version of these formulas, that is, a version that holds for either a general coordinate system or a curved spacetime, see p. 74 of Ref. \cite{Synge1}). The functions $k_1$, $k_2$ and $k_3$ are known as first, second and third curvatures, respectively. The curvature $k_1$ measures how rapidly the curve pulls away from the tangent line at $s$, while $k_2$ and $k_3$  measure, respectively, how rapidly the curve pulls away from the plane formed by $e_{(0)}, e_{(1)}$ and from the hyperplane formed by $e_{(0)}, e_{(1)}, e_{(2)}$ at $s$ (for more details, see Ref. \cite{Formiga:2006da}). 

It is important to note that when $k_1$ is zero, only $e_{(0)}$ is defined by the previous formulas. To keep the geometrical meaning of $k_2$ and $k_3$, we have to set them equal to zero. In this case, the vectors $e_{(i)}$ must be constant. The same happens with $k_3$ if $k_2$ vanishes.  However, if we are not worried about the meaning of $k_i$, we can choose the vectors that are not fixed by these formulas as we wish; of course, they have to satisfy the requirements to be a tetrad basis.

\section{The Proper Coordinate System of an accelerated observer \label{21102012s2}}

In this section, I consider the worldline of two distinct observers and choose a frame that is attached to one of them to construct a vector field globally defined in the Minkowski spacetime. After that, I impose the condition needed to ensure that we are using the proper coordinate system of the chosen accelerated observer.

\begin{figure}[h]
\includegraphics[scale=1.5]{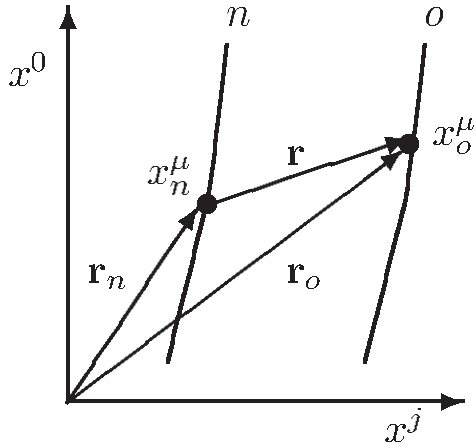}
\caption{This figure shows two observers at $x^{\mu}_n$, $x^{\mu}_o$ and their respective worldline, $n$ and $o$. The 4-vectors ${\bf r}_{n}$ and ${\bf r}_{o}$ represents the events $x^{\mu}_n$ and $x^{\mu}_o$, respectively; the 4-vector ${\bf r}$ is the position of $x^{\mu}_o$ relative to $x^{\mu}_n$. Here, $x^j$ represents the axis $x$, $y$ and $z$. }
\label{figure1}
\end{figure}

To begin with, let two observers $n$ and $o$ describe the curves $x^{\mu}_n(s_n)$ and $x^{\mu}_o(s_o)$ in an inertial frame of reference $I$ (see figure \ref{figure1}). Now, let  $\Lambda(s_n)$ be a local Lorentz transformation from $I$ to another inertial frame that, in an instant $s_n/c$ (c is the speed of light), coincides with a noninertial frame $S$ attached to the observer $n$. In searching for the proper coordinate system of $n$, we want the following to hold:
\begin{equation}
(x^{\nu}_o-x^{\nu}_n)\Lambda^{0}_{\ \nu}(s_n)=0, \label{1582012b}
\end{equation}
that is, both events $x^{\mu}_n(s_n)$ and $x^{\mu}_o(s_o)$ are simultaneous in the frame $S$. In what comes next, it is more suitable to use a different approach. Instead of dealing with coordinates directly, I shall deal with vectors first; then, when necessary or convenient, I use coordinates. 

As figure \ref{figure1} suggests, the relation among the vectors $\n{r}_n$, $\n{r}_o$ and $\n{r}$ is
\begin{equation}
\n{r}=\n{r}_o-\n{r}_n. \label{1582012a}
\end{equation}  

Let $\tilde{e}_A(s_n)$ be the frame that is attached to the observer $n$ at an instant $s_n/c$ (the frame $S$), not necessarily the Frenet-Serret tetrad. From this definition, one defines the co-frame $\tilde{e}^A(s_n)$ (also called dual basis) through $\tilde{e}^A(\tilde{e}_B)=\delta^A_B$. The components of $\tilde{e}_A$ in a coordinate basis $\tilde{\partial}_{\mu}$ will be denoted by $\tilde{e}_A^{\ \ \mu}$, while the ones for the co-frame will be denoted by $\tilde{e}^A_{\ \ \mu}$. 

In the definition above, both the frame and the co-frame are defined along the worldline of the observer $n$. Nonetheless, we can parallel transport them to  an arbitrary point so that we construct a vector and a co-frame field defined everywhere. Let us denote the transported vector by $e_A$ and the co-frame by $e^A$, where their components are also defined without the ``tilde''. It is well known that neither Cartesian basis nor the components of the vectors written in this basis change under parallel transport in the Minkowski spacetime. Therefore, if we take $\tilde{\partial}_{\mu}$ as being the Cartesian basis, we can identify $\tilde{\partial}_{\mu}$ with $\partial_{\mu}$ and $\tilde{e}_A^{\ \ \mu}$ with $e_A^{\ \ \mu}$; the same also holds for the co-frame, which is written in terms of $dx^{\mu}$. Since we shall deal only with Cartesian basis, the ``tilde'' will be omitted from now on. As a result, we have both the vector field $e_A$ and the one-form field $e^A$  globally defined in the Minkowski spacetime and having the same form at each point of this manifold.

From the arguments above, we can write $\n{r}_o=x^{\mu}_o\partial_{\mu}=x^{\mu}_o e^A_{\ \ \mu}e_A$ and  $\n{r}_n=x^{\mu}_n\partial_{\mu}=x^{\mu}_n e^A_{\ \ \mu}e_A$, where it was used the identity $\partial_{\mu}=e^A_{\ \ \mu} e_A$. Using these expressions and the definition $\n{r}= r^A e_A$ in Eq. (\ref{1582012a}), we may write

\begin{equation}
r^A=(x^{\mu}_o-x^{\mu}_n) e^A_{\ \ \mu}. \label{1582012c}
\end{equation}
In this approach, the equivalent version of (\ref{1582012b}) is
\begin{equation}
r^{(0)}=(x^{\mu}_o-x^{\mu}_n) e^{(0)}_{\ \ \mu}=0, \label{1582012d}
\end{equation}
where $e^A_{\ \ \mu}$ plays the role of the local Lorentz transformation, and it was assumed that $e_{(0)}^{\ \ \mu}\equiv d x^{\mu}_n/ds_n$. Recall that the components of the parallel-transported vectors do not change because they are written in terms of the Cartesian basis. Therefore, the vector field $e_{(0)}^{\ \ \mu}= d x^{\mu}_n/ds_n$ has this form everywhere.

By using (\ref{1582012d}), we can invert (\ref{1582012c}) to get
\begin{equation}
x^{\nu}_o=x^{\nu}_n+r^{(j)} e_{(j)}^{\ \ \nu}. \label{1582012e}
\end{equation}
The reason why I am writing this is because $x^{\nu}_n$, and consequently $e_{(j)}^{\ \ \nu}$, is supposed to be known and we want to know how the observer $o$ is described in the frame $S$; the pair $(s_n,r^{(i)})$ will represent the observer $o$ in $S$. 

From (\ref{1582012e}), we can create a set of static observers by taking $r^{(j)}$ constant.  For instance, if we take $r^{(i)}=0$, we have the observer $n$; for any other value, we have another observer who is at a fixed proper distance from the observer $n$.  However, for a general observer $o$, not necessarily at a fixed distance from $n$, we can see $x^{\nu}_o$ in (\ref{1582012e}) as a function of $\tau$ and $r^{j}$, where $\tau=\tau_n=s_n/c$. Therefore, differentiation of (\ref{1582012e}) leads to
\begin{equation}
dx^{\nu}_o=\left( \frac{d x^{\nu}_n}{d\tau}+r^{(j)} \frac{d e_{(j)}^{\ \ \nu}}{d\tau} \right)d\tau+ e_{(j)}^{\ \ \nu} dr^{(j)}. \label{1782012a}
\end{equation}

By using $ds^2=\eta_{\mu \nu} dx^{\mu}_o dx^{\nu}_o$, we get
\begin{widetext}
\begin{eqnarray}
ds^2=\left[c^2+2cr^{(i)} e_{(0)\mu}\frac{d e_{(i)}^{\ \ \mu}}{d\tau}+r^{(i)}r^{(j)} \eta_{\mu \nu} \frac{d e_{(i)}^{\ \ \nu}}{d\tau} \frac{d e_{(j)}^{\ \ \mu}}{d\tau}   \right]d\tau^2 +2r^{(i)} e_{(j)\nu}\frac{d e_{(i)}^{\ \ \nu}}{d \tau} d\tau dr^{(j)}
\nonumber \\
+\eta_{(i)(j)} dr^{(i)} dr^{(j)}, \label{1782012b}
\end{eqnarray}
\end{widetext}
where $r^{(1)}$, $r^{(2)}$, and $r^{(3)}$ are such that
\begin{equation}
c^2+2cr^{(i)} e_{(0)\mu}\frac{d e_{(i)}^{\ \ \mu}}{d\tau}+r^{(i)}r^{(j)} \eta_{\mu \nu} \frac{d e_{(i)}^{\ \ \nu}}{d\tau} \frac{d e_{(j)}^{\ \ \mu}}{d\tau}>0. \label{1782012h}
\end{equation}
It is important to keep in mind that (\ref{1782012b}) holds only if $e_A$ and $e^A$ are written in terms of the Cartesian basis $\partial_{ct}$, $\partial_x$, $\partial_y$ and $\partial_z$.

It is clear in (\ref{1782012b}) that $\tau$ and $r^{(i)}$ are the proper coordinates of $n$. To see this, we just need to set $\tau=constant$ and verify that $ds^2=\eta_{(i)(j)} dr^{(i)} dr^{(j)}$. Of course, by definition, $ds_n=cd\tau$  ($r^{(i)}=0$). However, it is interesting to note that although the proper distances of both $n$ and $o$ are the same, the proper time of an observer $o$ at a fixed proper distance from the observe  $n$ is not $\tau$, but rather
\begin{eqnarray}
c^2d\tau^2_o=\biggl[c^2+2cr^{(i)} e_{(0)\mu}\frac{d e_{(i)}^{\ \ \mu}}{d\tau}
+r^{(i)}r^{(j)} \eta_{\mu \nu} \frac{d e_{(i)}^{\ \ \nu}}{d\tau} \frac{d e_{(j)}^{\ \ \mu}}{d\tau}   \biggr]d\tau^2. \label{1782012c}
\end{eqnarray}

\subsection{The metric and the curvatures of the observer's curve}
Here, I write the line element (\ref{1782012b}) in terms of the curvatures of the curve described by the observer $n$, although the interpretation of $k_i$ as the curvatures of this observer's worldline cannot always be true, as we shall see in Sec. \ref{24102012s1}.

By choosing the basis $e_A$ to be the parallel transported version of Frenet-Serret basis (see Eqs. (\ref{22102012a})-(\ref{22102012d})), the line element (\ref{1782012b}) can be written as
\begin{widetext}
\begin{eqnarray}
ds^2=\left[(1+k_1r^{(1)})^2-(k_3^2+k_2^2)(r^{(2)})^2-( k_2r^{(1)}-k_3r^{(3)}   )^2   \right]c^2d\tau^2 +2\Bigl[  -k_2\delta_{j2}r^{(1)}
\nonumber \\
+(k_2\delta_{j1}-k_3\delta_{j3})r^{(2)}+k_3\delta_{j2}r^{(3)}  \Bigr]cd\tau dr^{(j)}-\delta_{ij} dr^{(i)}dr^{(j)}, \label{1782012d}
\end{eqnarray}
\end{widetext}
where
\begin{equation}
(1+k_1r^{(1)})^2-(k_3^2+k_2^2)(r^{(2)})^2-( k_2r^{(1)}-k_3r^{(3)}   )^2>0. \label{1782012i}
\end{equation}
The Frenet-Serret tetrad is defined only along the curve of the observer $n$. Nonetheless, as described at the beginning of this section, we can use the parallel transport to create a vector field that is defined everywhere and has the same form as that of the one defined along $x^{\mu}_n$.

In $2+1$ dimensions, we have $k_3=0$ (see \cite{Formiga:2006da} for more details). Hence, the line element (\ref{1782012d}) reduces to
\begin{widetext}
\begin{eqnarray}
ds^2=\left\{(1+k_1r^{(1)})^2-k_2^2\left[ (r^{(2)})^2 +(r^{(1)})^2 \right]   \right\}c^2d\tau^2 
\nonumber \\
+2k_2 \Bigl[  \delta_{j1}r^{(2)} -\delta_{j2}r^{(1)} \Bigr]cd\tau dr^{(j)}-\delta_{ij} dr^{(i)}dr^{(j)}. \label{1782012e}
\end{eqnarray}
\end{widetext}
In the next section, I use (\ref{1782012e}) to obtain the line element for the Rindler and the rotating observers.

\section{The rigid motion problem}
In this section I show how we can use Eqs. (\ref{1582012e}) and (\ref{1782012d}) to create a systematic way to construct a rigid motion in the sense of Ref. \cite{Rosen1947}.

The definition of rigid motion in special relativity was first given by Born \cite{Born1909}. This definition corresponds to a very strong constraint and, to relax it, one may use  the following definition \cite{Rosen1947}:
\begin{equation}
p_{\mu \nu} \equiv \frac{1}{2} \left(u_{\mu;\nu}+u_{\nu;\mu}-u_{\mu;\alpha}u^{\alpha} u_{\nu}-u_{\nu;\alpha}u^{\alpha} u_{\mu} \right)=0, \label{8032013e}
\end{equation}
where the semicolon denotes covariant differentiation, and $u^{\mu}=d x^{\mu}/ds$.

It is clear from Eq. (\ref{8032013e}) that not all kinds of motion are allowed, which can be considered as an unsatisfactory fact because one would rather have a definition of rigidity that was independent of the motion, as in a Euclidean space. However, it seems impossible to have such a definition.

\subsection{Allowed motions}

Let us now consider the observers that are characterized by the constant values of $\xi \equiv r^{(1)}$, $\chi \equiv r^{(2)}$, and $\lambda \equiv r^{(3)}$, which do not impose any restriction on the possible motions of the observer $n$. In these coordinates, the observer $o$ is described by $(\tau, \xi, \chi, \lambda)$. The $4$-velocity of this observer is
\begin{equation}
u^{\mu}=\frac{d\tau}{ds_o} \delta^{\mu}_{\ 0}=\frac{1}{c} f^{-1/2} \delta^{\mu}_{\ 0}, \label{23062013b}
\end{equation}
where I have used (\ref{1782012d}), omitted the ``$o$'' in $u^{\mu}$ and defined $f\equiv (1+k_1\xi)^2-(k_3^2+k_2^2)\chi^2-( k_2\xi-k_3\lambda   )^2$. The covariant component of the $4$-velocity is
\begin{equation}
u_{\nu}=\frac{1}{c}f^{-1/2}g_{0\nu} \label{23062013m}
\end{equation}
From now on, I shall use only the letters $\xi$, $\chi$, and $\lambda$ to represent the coordinates $r^{(i)}$.

The $4$-velocity (\ref{23062013m}) allows us to write the tensor $p_{\mu \nu}$ in the following convenient form:
\begin{eqnarray}
p_{\mu \nu}=\frac{1}{2}\Bigl[  u_{\mu,\nu}+u_{\nu,\mu} -\frac{1}{c}f^{-1/2} \left( u_{\mu,0}u_{\nu}+u_{\nu,0}u_{\mu}  \right) -\frac{2}{c}f^{-1/2}\Gamma_{0\mu \nu}
\nonumber \\
+\frac{1}{c^2}f^{-1}\left( \Gamma_{00\mu}u_{\nu}+\Gamma_{00\nu}u_{\mu}  \right)  \Bigr], \label{23062013n}
\end{eqnarray}
where $\Gamma_{\lambda \mu \nu}\equiv (1/2)(g_{\mu \lambda,\nu}+g_{\nu \lambda,\mu}-g_{\mu \nu,\lambda})$ is the Christoffel symbol of the first kind.

The condition $p_{\mu \nu}=0$ does not hold for an arbitrary motion. Hence, it is important to known under what conditions the observers $o$ are rigid. The following theorem establishes necessary and sufficient conditions for this to happen.

\begin{theorem} \label{23062013a}
Let $k_1$, $k_2$, and $k_3$ be the curvatures of the curve described by the observer ``$n$''. The set of observers ``$o$'' defined by the constant values of $\xi$, $\chi$, and $\lambda $ will represent a rigid motion in the sense of Eq. (\ref{8032013e}) if and only if the curvatures $k_2$ and $k_3$ are constant. In addition, for a nonconstant $k_1$, these observers are rigid if and only if both $k_2$ and $k_3$ vanish.
\end{theorem}

The proof goes as follows. If Eq. (\ref{8032013e}) holds, then from the components $p^{11}$ and $p^{33}$ we  respectively have
\begin{equation}
k_2 \left[ \dot{k}_2+\left(2k_1\dot{k}_2-k_2\dot{k}_1  \right)\xi +\left(k_1^2\dot{k}_2-k_2k_1 \dot{k}_1 \right)\xi^2 \right]=0, \label{23062013c}
\end{equation}
\begin{equation}
-k_3 \left[ \dot{k}_3+k_2^2\dot{k}_3-k_2k_3\dot{k}_2+\left(2k_1\dot{k}_3-k_3\dot{k}_1  \right)\xi +\left(k_1^2\dot{k}_3-k_3k_1 \dot{k}_1 \right)\xi^2 +\left(k_3k_2\dot{k}_2-k_2^2\dot{k}_3 \right)\chi^2 \right]=0, \label{23062013d}
\end{equation}
where the overdot stands for derivative with respect to $\tau$. By taking into account  that the coordinates are arbitrary and assuming $k_2 \neq 0$ in Eq. (\ref{23062013c}), we arrive at
\begin{eqnarray}
\dot{k}_2=0, \label{23062013e}
\\
2k_1\dot{k}_2-k_2\dot{k}_1=0, \label{23062013f}
\\
k_1^2\dot{k}_2-k_2k_1 \dot{k}_1=0. \label{23062013g}
\end{eqnarray}
From Eq. (\ref{23062013d}) and the assumption $k_3 \neq 0$, we obtain
\begin{eqnarray}
\dot{k}_3+k_2^2\dot{k}_3-k_2k_3\dot{k}_2=0, \label{23062013h}
\\
2k_1\dot{k}_3-k_3\dot{k}_1=0, \label{23062013i}
\\
k_1^2\dot{k}_3-k_3k_1 \dot{k}_1=0, \label{23062013j}
\\
k_3k_2\dot{k}_2-k_2^2\dot{k}_3=0. \label{23062013l}
\end{eqnarray}
It is clear in Eq. (\ref{23062013e}) that $k_2$ must be constant. Besides, from Eqs. (\ref{23062013h}) and (\ref{23062013l}), one easily prove that $\dot{k}_3$ must vanish independently of $k_2$. In turn, from Eqs. (\ref{23062013f}) and (\ref{23062013i}) we see that if $k_1$ is not constant, then $k_2$ and $k_3$ must be zero.

Let us now see that, for constant curvatures, Eq. (\ref{8032013e}) is identically satisfied. In this case, the components of the Christoffel symbol that are of our interest are
\begin{eqnarray}
\Gamma_{001}=c^2\left[ k_1+k_2k_3\lambda +(k_1^2-k_2^2)\xi \right],\quad \Gamma_{002}=-c^2(k_2^2+k_3^2)\chi,
\nonumber \\
 \Gamma_{003}=c^2(k_2k_3\xi-k_3^2\lambda), \quad \Gamma_{012}=-ck_2/2.
\end{eqnarray}
By using Eqs. (\ref{1782012d}), (\ref{23062013b}), (\ref{23062013m}), and the expressions above in Eq. (\ref{23062013n}), one can check that $p_{\mu \nu}=0$.

To finish the proof of theorem \ref{23062013a}, we just need to verify that Eq. (\ref{8032013e}) holds for an arbitrary $k_1$ as long as $k_2$ and $k_3$ vanish. For this case, we have
\begin{equation}
f=(1+k_1\xi)^2, \quad   u_{\mu}=c(1+k_1\xi)\delta^0_{\ \mu}, \quad \Gamma_{000}=c^2( \dot{k}_1+k_1\dot{k}_1\xi )\xi, \quad \Gamma_{001}=c^2(1+k_1\xi)k_1.
\end{equation}
From these expressions and Eq. (\ref{23062013n}), it is straightforward to check that $p_{\mu \nu}$ vanishes, which finishes our proof of the theorem \ref{23062013a}.

We can use the theorem \ref{23062013a} and the accelerated observers that are static in the coordinates $\xi$, $\chi$, and $\lambda$ to obtain a particular rigid motion. Examples of how this can be done are given in the next section.

\section{Applications \label{21102012s3}}
To exemplify the application of the observers considered in the previous section, I obtain the Rindler observers, the rotating ones, and create a new set of rigid observers.

\subsection{Rindler Observers}
In $1+1$,  an observer whose 4-acceleration $a$ is constant can described, for certain initial conditions, by  \cite{Formiga:2006ur}
\begin{eqnarray}
x^0=\frac{c^2}{a}\sinh\left( \frac{a\tau}{c}\right), \label{8032013a} \\
x^1=\frac{c^2}{a}\cosh\left( \frac{a\tau}{c}\right). \label{8032013b}
\end{eqnarray}
We can use this observer as the observer $n$ in order to get the observer $o$ and, then, construct a ``rigid rod'' that is accelerated with a constant $4$-acceleration. 

By using Eqs. (\ref{8032013a}) and (\ref{8032013b}) into (\ref{22102012a})-(\ref{22102012d}), we get $k_1=a/c^2$ and $k_2=k_3=0$. Besides, Eq. (\ref{1582012e}) becomes  
\begin{eqnarray}
x^0=\left(\frac{c^2+\xi a}{a} \right)\sinh\left( \frac{a\tau}{c}\right), \label{8032013c}\\
x^1=\left(\frac{c^2+\xi a}{a} \right)\cosh\left( \frac{a\tau}{c}\right), \label{8032013d}
\end{eqnarray}
which defines the so-called Rindler observers (see Ref. \cite{Formiga:2006ur} for more details). 

It is easy to see that the line element (\ref{1782012e}) reduces to the well-known expression
\begin{equation}
ds^2=(1+\frac{a}{c^2} \xi)^2c^2d\tau^2-d\xi^2 \label{1882012a}
\end{equation}
 and the condition (\ref{1782012i}) implies $ \xi \in (-c^2/a,\infty)$.

Rindler observers can clearly mimic a rigid rod since the curvatures of the observer $n$ are constant. It is straightforward to verify that Eqs. (\ref{8032013c}) and (\ref{8032013d}) yield 
\begin{eqnarray}
u^0=x^1/\sqrt{(x^1)^2-(x^0)^2},\\
u^1=x^0/\sqrt{(x^1)^2-(x^0)^2},
\end{eqnarray}
which satisfy Eq.(\ref{8032013e}). 

\subsection{Rotating observers \label{24102012s1}}
Let an observer $n$ that is rotating with a constant angular velocity $\omega $ and at a distance $R$ from the origin of a inertial frame $I$ have the coordinates
\begin{equation}
x^{0}_n=ct,\quad x^{1}_n=R\cos\omega t,\quad x^{2}_n=R\sin\omega t. \label{1882012d}
\end{equation}
By using the Frenet-Serret basis, we obtain
\begin{eqnarray}
e_{(0)}^{\ \ \mu}=\gamma (1,-\frac{\omega R}{c}\sin\omega t,\frac{\omega R}{c}\cos\omega t,0), \label{1882012e}\\
e_{(1)}^{\ \ \mu}=(0,-\cos\omega t,-\sin\omega t,0), \\
e_{(2)}^{\ \ \mu}=\gamma ( -\frac{\omega R}{c}, \sin\omega t, -\cos\omega t,0 ), \label{1882012f}\\
e_{(3)}^{\ \ \mu}=(0,0,0,1), \label{1882012g}
\end{eqnarray}
where $\gamma=1/\sqrt{1-\omega^2R^2/c^2}$. From Eq. (\ref{1582012e}), we get
\begin{eqnarray}
x^0_o=ct-\omega R \gamma \chi/c, \label{25062013a}\\
x^1_o=(R-\xi)\cos\omega t+\chi\gamma\sin\omega t, \label{25062013b}\\
x^2_o=(R-\xi)\sin\omega t-\chi\gamma\cos\omega t. \label{25062013c}
\end{eqnarray}
These are the coordinates of $o$ in the frame $I$.

The curvatures of the observer $n$ are $k_3=0$, and 
\begin{eqnarray}
k_1=\gamma^2 \frac{\omega^2 }{c^2}R, \label{1882012h}\\
k_2=\gamma^2 \frac{\omega }{c}. \label{1882012i}
\end{eqnarray}
These curvatures are clearly constant, which allows us to use this observer $n$ to construct a rigid set of observers which rotates with it. We construct this set by taking $\xi$ and $\chi$ constant in the coordinates (\ref{25062013a})-(\ref{25062013c}).

The substitution of $k_1$ and $k_2$ into (\ref{1782012e}) gives
\begin{widetext}
\begin{eqnarray}
ds^2=\left[ (1+\gamma^2\frac{\omega^2 R}{c^2}\xi )^2-\gamma^4\frac{\omega^2}{c^2}( \xi^2+\chi^2  )  \right]c^2d\tau^2+2\gamma^2\omega ( \chi d\xi -\xi d\chi  )d\tau -d\xi^2 -d\chi^2. \label{1782012f}
\end{eqnarray}
\end{widetext}

For simplicity, let us set $R=0$. In this case, we have
\begin{widetext}
\begin{eqnarray}
ds^2=\left[ 1-\frac{\omega^2}{c^2}( \xi^2+\chi^2  )  \right]c^2d\tau^2+2\omega ( \chi d\xi -\xi d\chi  )d\tau -d\xi^2 -d\chi^2, \label{1782012g}
\end{eqnarray}
\end{widetext}
where condition (\ref{1782012i}) leads to $\sqrt{\xi^2+\chi^2}< c/\omega$. This is the same line element of the rigid disk in Ref. \cite{Berenda:1942zz}. 

Here, we have to be very careful with the meaning of $k_i$. When $k_1$ is zero the Serret-Frenet formulas do not determine $k_2$, $k_3$, $e_{(2)}$ and $e_{(3)}$, as pointed out before. This is exactly the case when one sets $R=0$ in Eq. (\ref{1882012d}) before evaluate the basis. On the other hand, when we perform the calculations first and then take $R=0$, we obtain $k_1=k_3=0$ and $k_2=\omega /c$, and the basis remains well defined. The problem in this case is that $k_i$ cannot be interpreted as the curvatures of a curve, since a curve which does not curve ($k_1=0$) cannot twist ($k_2 \neq 0$). For our purpose this is irrelevant, since we do not need $k_i$ to be curvatures. But now, we have the question: ``why don't we take the inertial frame, since it also satisfies Frenet-Serret formulas?'' The answer is simple: the rotating observers who are not at the origin ($\xi\ or \ \chi \neq 0$) must be at rest with respect to the chosen frame so that they keep their rigidity in this frame.  To understand better, consider the following. If we have just one particle at the origin, then we have two types of frame that the particle can be at rest: a frame that rotates around its origin, and a frame that does not rotate at all. However, if we have a rigid disk made of particles that are at rest with respect to a certain frame, there will  be only one frame satisfying this condition: the one which rotates together with the particles. Therefore, we have to choose that frame for the rotating disk.

\subsection{A new set of accelerated observers and its rigid motion \label{28102012s1}}
\begin{figure}[h]
\includegraphics[scale=0.5,angle=-90]{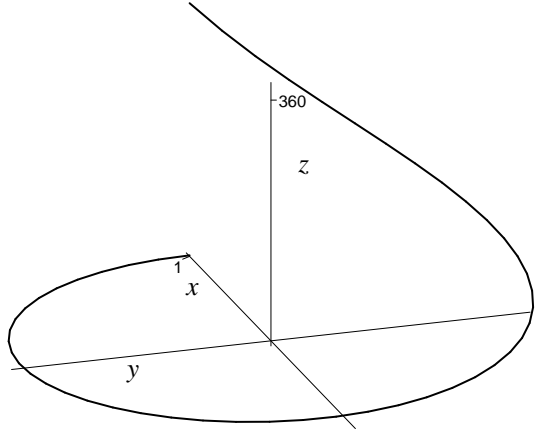}
\caption{In this figure, the projection of the curve (\ref{28102012f}) on $x,y,z$ is shown for $\sqrt{3}c^2/a=1$ and $\theta$ starting from $0$.}
\label{30102012a}
\end{figure}

 Let the observer $n$ describe the following path in the inertial frame $I$:
\begin{equation}
x_n^{\mu}=\sqrt{3}\left(\frac{\sqrt{2}c^2}{a}\sinh\theta,\frac{c^2}{a}\cos\theta,\frac{c^2}{a}\sin\theta,\frac{\sqrt{2}c^2}{a}\cosh\theta \right), \label{28102012f}
\end{equation}
where $\theta=as/(\sqrt{3}c^2)$ ($s$ is the arc length), and $a$ is the observer's 4-acceleration. This observer not only rotates around the $z$-axis but also translates along it (see figure \ref{30102012a}).

From the coordinates $x_n^{1}$, $x_n^{2}$ and the observer's proper time, we can see that the observer (\ref{28102012f}) rotates around the $z$-axis with a constant angular speed from his point of view. By equating $\theta$ to $2m\pi$ ($m=0,1,2...$), we get $c\tau_m=s_m=2m\pi \sqrt{3}c^2/a$, which yields the period $T=2\sqrt{3}c\pi/a$. However, from the point of view of the observer who is at rest in the $I$ frame, this is not a periodic rotation because the function ``$\sinh$'' is not periodic ($t_{m+1}-t_m$ depends on $m$). 

By using the Frenet-Serret formulas, one obtains the tetrad
\begin{eqnarray}
e_{(0)}^{\ \ \mu}=\left(\sqrt{2}\cosh\theta,-\sin\theta,\cos\theta,\sqrt{2}\sinh\theta \right), \label{28102012a}\\
e_{(1)}^{\ \ \mu}=\frac{1}{\sqrt{3}}\left( \sqrt{2}\sinh\theta,-\cos\theta,-\sin\theta,\sqrt{2}\cosh\theta \right), \label{28102012b}\\
e_{(2)}^{\ \ \mu}=\frac{1}{\sqrt{2}} \left( -\sqrt{2}\cosh\theta,2\sin\theta,-2\cos\theta,-\sqrt{2}\sinh\theta \right), \label{28102012c}\\
e_{(3)}^{\ \ \mu}=\frac{1}{\sqrt{3}} \left(\sinh\theta,\frac{2}{\sqrt{2}}\cos\theta,\frac{2}{\sqrt{2}}\sin\theta,\cosh\theta \right), \label{28102012d}
\end{eqnarray}
and the curvatures
\begin{equation}
k_1=\frac{a}{c^2},\quad k_2=\frac{2\sqrt{2}}{3}\frac{a}{c^2},\quad k_3=\frac{1}{3}\frac{a}{c^2}, \label{28102012e}
\end{equation}
which are constant.

The substitution of (\ref{28102012e}) into (\ref{1782012d}) yields
\begin{eqnarray}
ds^2=\left[\left(1+\frac{a\xi}{c^2} \right)^2-\frac{a^2}{9c^4} \left( 2\sqrt{2}\xi-\lambda \right)^2-\frac{a^2}{c^4}\chi^2  \right]c^2d\tau^2
\nonumber \\
+\frac{2a}{3c}\left[  2\sqrt{2}( \chi d\xi-\xi d\chi  ) +\lambda d\chi -\chi d\lambda \right]d\tau
\nonumber \\
-d\xi^2-d\chi^2-d\lambda^2.
\end{eqnarray}
Remember that $\xi=r^{(1)}$, $\chi=r^{(2)}$ and $\lambda= r^{(3)}$. From (\ref{1582012e}), we get
\begin{eqnarray}
x_{o}^{0}=A\sinh\theta -\sqrt{2}C\cosh\theta, \quad x_{o}^{1}=B\cos\theta +2C\sin\theta,
\nonumber \\
x_{o}^{2}=B\sin\theta -2C\cos\theta, \quad x_{o}^{3}=A\cosh\theta -\sqrt{2}C\sinh\theta, \label{10032013a}
\end{eqnarray}
where
\begin{equation}
A=\sqrt{6} \left(  \frac{c^2}{a}+\frac{\xi}{3}+\frac{\lambda}{3\sqrt{2}} \right), \quad B=\sqrt{3} \left(  \frac{c^2}{a}-\frac{\xi}{3}+\frac{2\lambda}{3\sqrt{2}} \right), \quad C=\frac{\chi}{\sqrt{2}}.
\end{equation}
One can easily verify that this curve satisfies $(x_o^0)^2-(x_o^1)^2-(x_o^2)^2-(x_o^3)^2=-(A^2+B^2+2C^2)$. 

By taking $\xi$, $\chi$, and $\lambda$ as constant, we obtain a set of observers which represent a rigid motion (see theorem \ref{23062013a}). Perhaps, these observers may be interpreted as a particular case of a solid cylinder that rotates around its axis and, at the same time, is accelerated along it.  

To double check that these observers satisfy Eq. (\ref{8032013e}), we can write the $4$-velocity in terms of the Cartesian coordinate $x^{\mu}$, where I have dropped the ``$o$'', and substitute the result into Eq. (\ref{8032013e}). By doing that, we arrive at $u^0=z/\alpha$, $u^1=-y/\alpha$, $u^2=x/\alpha$ and $u^3=w/\alpha$, where I have used $\alpha=\sqrt{z^2-y^2-x^2-w^2}$ and $w\equiv x^0$, $x\equiv x^1$, $y\equiv x^2$, $z\equiv x^3$. One can easily check that this $4$-velocity satisfies Eq. (\ref{8032013e}).

\section{Final Remarks}
The proper coordinate system used in (\ref{1782012b}) belongs to the observer $n$ and, in this sense, this observer is privileged. This is not a strange fact because, in general, each observer has a different 3-velocity with respect to the frame $I$. For instance, the magnitude of the 3-velocity of a Rindler observer can be shown to be
\begin{equation}
V=\frac{at}{  \sqrt{(1+a\xi/c^2)^2+a^2t^2/c^2}  },
\end{equation}
where $t$ and $x$ are the Cartesian coordinates used by the inertial observers in $I$. This velocity clearly depends on the position of the Rindler observers, that is, on $\xi$. Hence, if we choose to write the metric in terms of proper distances and a proper time for a set of accelerated observers, in general, we will have to choose the proper coordinate system of one of them. For a coordinate system that does not privilege any of the Rindler observers, see Ref. \cite{daSilva:2007zz}

The advantage of using the Frenet-Serret formalism lies in the fact that, with the help of theorem \ref{23062013a}, it allows for a systematic and easy way to construct any rigid set of observers. One of the main advantages is that we do not need to know whether the curve of the observer $n$ (the one used to generate the whole set) is inside a plane or in a hyperplane (or even in the whole spacetime), since the calculations already give the simplifications needed when the curve lies either in a plane or in a hyperplane. For example, when the curvature $k_3$ vanishes, all nontrivial terms with $r^{(3)}$ disappear in the metric (\ref{1782012d}). The same goes for both $r^{(2)}$ and $r^{(3)}$ when $k_2=k_3=0$.

%\begin{acknowledgments}
%\end{acknowledgments}

%\bibliography{bib}

%

\end{document}